\documentclass[11pt]{article}

\usepackage{arxiv}

\usepackage[utf8]{inputenc} % allow utf-8 input
\usepackage[T1]{fontenc}    % use 8-bit T1 fonts
\usepackage{hyperref}       % hyperlinks
\usepackage{url}            % simple URL typesetting
\usepackage{booktabs}       % professional-quality tables
\usepackage{amsfonts}       % blackboard math symbols
\usepackage{amsmath}
\usepackage{amssymb}
\usepackage{nicefrac}       % compact symbols for 1/2, etc.
\usepackage{microtype}      % microtypography
\usepackage{lipsum}		% Can be removed after putting your text content
\usepackage{graphicx}
\usepackage{natbib}
\usepackage{doi}
\usepackage{xcolor}
\usepackage{listings}
\definecolor{commentgreen}{RGB}{2,112,10}
\definecolor{codegreen}{rgb}{0,0.6,0}
\definecolor{codegray}{rgb}{0.5,0.5,0.5}
\definecolor{codepurple}{rgb}{0.58,0,0.82}
\definecolor{backcolour}{rgb}{0.95,0.95,0.92}

\lstdefinelanguage{ASM}{
    morekeywords={sin, time, rateOf, uniform, normal,
       maximize, constraint, model, end, notes},
    sensitive=false, % keywords are not case-sensitive
    morecomment=[l]{//}, % l is for line comment
    morecomment=[s]{/*}{*/}, % s is for start and end delimiter
    morestring=[b]" % defines that strings are enclosed in double quotes
} % 

\lstdefinestyle{mystyle}{
    language=ASM,
    commentstyle=\color{codegreen},
    keywordstyle=\color{magenta},
    numberstyle=\tiny\color{codegray},
    stringstyle=\color{codepurple},
    basicstyle=\ttfamily,%\footnotesize,
    breakatwhitespace=false,         
    breaklines=true,                 
    captionpos=b,                    
    keepspaces=true,  
    frame=single,               
    numbersep=5pt,                  
    showspaces=false,                
    showstringspaces=false,
    showtabs=false, 
    columns=fullflexible,
    tabsize=2
}

\title{An Update to the SBML Human-Readable Antimony Language}

\author{\href{https://orcid.org/0000-0001-7002-6386}{\includegraphics[scale=0.06]{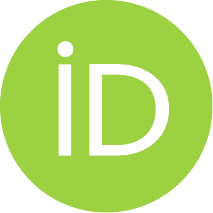}\hspace{1mm}Lucian Smith}\\
   Department of Bioengineering\\
   University of Washington\\
   Seattle, WA 98195-5061\\
   \texttt{lpsmith@uw.edu }\\
   \AND \href{https://orcid.org/0000-0002-3659-6817}{\includegraphics[scale=0.06]{orcid.pdf}\hspace{1mm}Herbert M.~Sauro}\\
	Department of Bioengineering\\
	University of Washington\\
	Seattle, WA 98195-5061 \\
	\texttt{hsauro@uw.edu} }

\renewcommand{\shorttitle}{Antimony Update}

\hypersetup{
pdftitle={Antimony Update},
pdfsubject={systems biology},
pdfauthor={Herbert M.~Sauro, Lucian Smith},
pdfkeywords={Antimony, SBML, Systems Biology},
colorlinks,
linkcolor=blue,
citecolor=blue,
urlcolor=blue
}

\begin{document}
\maketitle

\begin{abstract}
Antimony is a high-level, human-readable text-based language designed for defining and sharing models in the systems biology community. It enables scientists to describe biochemical networks and systems using a simple and intuitive syntax. It allows users to easily create, modify, and distribute reproducible computational models. By allowing the concise representation of complex biological processes, Antimony enhances collaborative efforts, improves reproducibility, and accelerates the iterative development of models in systems biology. This paper provides an update to the Antimony language since it was introduced in 2009. In particular, we highlight new annotation features, support for flux balance analysis, a new {\tt rateOf} method, support for probability distributions and uncertainty, named stochiometries, and algebraic rules. Antimony is also now distributed as a C/C++ library, together with python and Julia bindings, as well as a JavaScript version for use within a web browser. Availability: \url{https://github.com/sys-bio/antimony}.
\end{abstract}

%Updates (from LS):
%* L3v2 new default target
%* SBML FBC added

% keywords can be removed
\keywords{Antimony \and SBML \and Systems Biology}

\section{Introduction}

Antimony is a high-level, human-readable text-based language designed for defining and sharing models in the systems biology community. We developed it as a human-readable format to complement the XML-based SBML\citep{hucka2003systems} standard. In general, Antimony is converted to SBML which can then be loaded into the many compatible SBML simulators such as libroadrunner~\citep{welsh2023libroadrunner}. Often simulators will convert the SBML into a set of differential equations which are then solved by the simulator to generate time course data. 

Antimony uses a minimum of punctuation characters which renders the text easier to read and understand. Antimony is implemented using C/C++~\citep{smith2009antimony} and Bison as the grammar parser. However, the distribution also includes Python bindings which can be installed using pip to make it easy to use from Python. Julia bindings are also available to users. It is also available via the Tellurium package~\citep{choi2018tellurium,medley2018tellurium} and more recently a JavaScript/WASM version was created which allows the Antimony language to be used on the web. The website tool makesbml uses the Javascript version~\citep{jardine2023makesbml} generated from emscripten~(\url{https://emscripten.org/}). There is also a pure Javascript parser for the Antimony web editor~(\url{https://github.com/sys-bio/AntimonyEditor})

Antimony supports SBML Level 3, version 2, and the following SBML packages: Hierarchical Model Composition~\citep{smith2015sbml}, Flux Balance Constraints~\citep{olivier2015systems} and Distributions~\citep{smith2020systems}.

In this article, we will describe some of the major updates that have been incorporated into Antimony since its inception.

\section{Examples}

In this section, we will show some examples of basic Antimony functionality that doesn't include the new features. The first example is a single chemical reaction whose rate is governed by simple mass-action kinetics:

\medskip
\begin{lstlisting}[caption={Antimony model of a single reaction}]
 // Define a single reaction with reactant A and product B and rate law k1*A
 A -> B; k1*A

 // Initialize values
 A = 5
 B = 0
 k1 = 0.1
\end{lstlisting}

Multiple reactions can be included in a model. For example, here is a simple branched pathway where the third reaction uses a reversible mass-action rate law.

\medskip
\begin{lstlisting}[caption={Antimony model of a simple branched pathway}]
 // Define three reactions forming a simple branch
 A -> B; k1*A
 B -> C; k2*B
 B -> D; k3*B - k4*C

 // Initialize values
 A = 5; B = 0; C = 0; D = 0;
 k1 = 0.1; k2 = 0.2; k3 = 0.15; k4 = 3.4
\end{lstlisting}

Multiple reaction/product reactions with non-unity stoichiometries are also easily specified as follows:

\medskip
\begin{lstlisting}[caption={Antimony model using non-unity stoichiometries}]
 // A reaction with multiple reactants
 A + B -> C; k1*A*B
 // A reaction with non-unity stoichiometry
 2 C -> D + E; k2*C

 // Initialize values
 A = 5; B = 0; C = 0; D = 0; E = 0; k1 = 0.1; k2 = 0.2; 
\end{lstlisting}

Named reactions can also be specified as follows. A simulator may use the reaction name to request values for the reaction rate:

\medskip
\begin{lstlisting}[caption={Antimony model using named reactions}]
 // Name a reaction 'Jl'.
 J1: A + B -> C; k1*A*B

 // Initialize values
 A = 5; B = 2; C = 0; k1 = 0.1; 
\end{lstlisting}

\pagebreak
Another example shows how external species or species that are determined by external forcing functions (known as assignment rules in SBML) can be described. External species (or boundary species in SBML terminology) are by default fixed. In practice, a simulator will not generate a differential equation for such species. 

\medskip
\begin{lstlisting}[caption={Antimony model using a fixed species and a forcing function.}]
 // A reaction with an external reactant, A, indicated with a dollar symbol.
 // Species A will be fixed in any model generated from this script
 $A -> B; k1*A
 
 // A reaction with a forcing function. In this case, the function
 // is a sine wave of a given frequency and amplitude. Note the use of
 // the ':=' assignment symbol to indicate this equation is part of the model
 // and will be executed continuously when the model is simulated
 X := 1 + amp*sin (time*freq)
 $X -> Y; k1*X 

 // Initialize values
 A = 1
 amp = 2.5; freq = 0.5; k1 = 0.1; 
\end{lstlisting}

Rates of change can also be set directly: $dS/dt$ in Antimony is represented as $S'$.  When declaring a rate of change for a species that appears in other reactions, that species must be declared as a boundary species, much like a species with a forcing function.

\medskip
\begin{lstlisting}[caption={Antimony model using explicit rates of change}]
 // Define a reaction.
 J1: $A + B -> C; k1*A*B

 // The 'A' species here is not consumed by reaction J1, and instead increases 
 //  at an autocatalytic rate
 $A' = k2 * A

 // Initialize values
 A = 5; B = 2; C = 0; k1 = 0.1; k2 = 0.01
\end{lstlisting}

Other useful features include the ability to define discrete events based on a boolean expression, the ability to build modular models that can be combined to create a larger model. The documentation for these as well as additional features can be found at~(\url{https://tellurium.readthedocs.io/en/latest/antimony.html}).

In the following sections, we will describe several significant enhancements that have been made to the Antimony language since its inception. 

\section{Metadata Extensions}

The BioModels repository~\cite{malik2020biomodels} contains a large number of curated models. In almost all cases, the curators at BioModels have added useful metadata to the model such as the author of the model, a link to the paper etc. Antimony has been extended to allow users to easily add metadata to a model or when retrieving a model in SBML the metadata is shown in the Antimony model. The following is an example of metadata being added to a model using Antimony:

%https://github.com/combine-org/combine-specifications/blob/main/specifications/qualifiers-1.1.md#biology-qualifiers

\begin{verbatim}
  model model_source "http://identifiers.org/biomodels.db/BIOMD0000000141"
  model publication "http://identifiers.org/pubmed/15484883"
  model biological_system "http://identifiers.org/go/GO:0019228"
  model taxon "http://identifiers.org/taxonomy/40674"
  model created "2007-07-16T09:41:14Z"
  model modified "2015-02-25T11:18:57Z"
  model creator1.givenName "John"
  model creator1.familyName "Smith"
  model creator1.organization "Oxford"
  model creator1.email "John@university.edu"
  model notes "This model describes how..."
\end{verbatim}

Note that when using the metadata feature, a model must be bracketed by {\tt model name()} and {\tt end}. A simple example below illustrates this requirement:

\medskip
\begin{lstlisting}[caption={Antimony model using a fixed species and a forcing function.}]
model mymodel()
 S1 -> S2; k1*S1

 model notes "This is a small model"
end
\end{lstlisting}

%  https://tellurium.readthedocs.io/en/latest/antimony.html#language-reference search for 'model_entity'

\section{rateOf}

{\tt rateOf} is a function call that can be used to retrieve the rate of change of a species or a rate rule. It can not be used to obtain the rate of change of a reaction rate or the left-hand side of an assignment rule. An example is given below which illustrates how to obtain the rate of change of a species {\tt S1} or the value of a differential equation, {\tt C' = 10*k3}. The {\tt rateOf} method can also be inserted into an equation, such as a reaction rate equation, assignment rules, etc. 

\medskip
\begin{lstlisting}[caption={Retrieving the rare of change of a species or rate rule.}]
  J0: S0 -> S1; k0*S0
  J1: S1 -> S2; k1*S1

  X1 := rateOf(S1)

  C' = 10*k3

  X2 := rateOf(C)
\end{lstlisting}

%Anything that is changing due to rate rule, species, 
%Can't do assignment rules, and can't do reaction rates

\section{Probability Distributions}

The latest version of Antimony allows a modeler to initialize values drawn from a given distribution function. For example, in the code below, {\tt S0} and {\tt S1} are drawn from a normal distribution with mean 0.5 and a standard deviation 0.2.  This can be useful for example when investigating how uncertainties propagate through a model. Repeated runs of such a model will result in variations in the outputs.

The distribution functions can also be used in the outcome of an event. For example, if an event is used to detect when a species passes a certain value, we can change a rate constant based on a distribution. In the example below, if {\tt S1} exceeds 10, we set the rate constant {\tt k0} based on a uniform distribution between 2.6 and 5.5. Note that distribution may only be used in discrete contexts (such as initializing a value or resetting a value from an event) and cannot be used in continuous contexts such as in reaction rates or assignment rules. 

\medskip
\begin{lstlisting}[caption={Initializing values based on a probability distribution.}]
  J0: S0 -> S1; k0*S0
  J1: S1 -> S2; k1*S1

  if S1 > 10:
     k0 = uniform(2.5, 5.5)

  S0 = normal(0.5, 0.2)
  S1 = normal(0.5, 0.2)
  S2 = 0
\end{lstlisting}

A variety of distributions are provided and are shown in the table below. Details of these distributions can be found in the SBML distribution documentation~\cite{smith2020systems}.

\begin{tabular}{l}
normal(mean, stddev) \\
normal(mean, stddev, min, max) \\
uniform(min, max) \\
bernoulli(prob) \\
binomial(nTrials, probabilityOfSuccess) \\
binomial(nTrials, probabilityOfSuccess, min, max) \\
cauchy(location, scale) \\
cauchy(location, scale, min, max) \\
chisquare(degreesOfFreedom) \\
chisquare(degreesOfFreedom, min, max) \\
exponential(rate) \\
exponential(rate, min, max) \\
gamma(shape, scale) \\
gamma(shape, scale, min, max) \\
laplace(location, scale) \\
laplace(location, scale, min, max) \\
lognormal(mean, stdev) \\
lognormal(mean, stdev, min, max) \\
poisson(rate) \\
poisson(rate, min, max) \\
rayleigh(scale) \\
rayleigh(scale, min, max)
\end{tabular}

%Can't use it during a simulation except in an event

%Useful for initial assignments

\section{Uncertainty}

The SBML ‘Distributions’ package introduced a variety of ways to store information about the uncertainty of model elements. Antimony has now been extended to store this same information through the following syntax:

\medskip
\begin{lstlisting}[caption={Uncertainty information attached to variables.}]
A.mean = x
A.stdev = x  (or A.standardDeviation = x)
A.coefficientOfVariation = x
A.kurtosis = x
A.median = x
A.mode = x
A.sampleSize = x
A.skewness = x
A.standardError = x
A.variance = x
A.confidenceInterval = {x, y}
A.credibleInterval = {x, y}
A.interquartileRange = {x,y}
A.range = {x,y}
A.distribution = function()
A.distribution is "http://uri"
A.externalParameter = x || {x,y} || function()
A.externalParameter is "http://uri"
\end{lstlisting}

In the above examples, {\tt A} may be any symbol in Antimony with mathematical meaning; {\tt x} and {\tt y} may be either a symbol or a value (i.e. {\tt A.mean=2.4; A.confidenceInterval=\{S1, 8.2\});} function() may be any mathematical formula; and "http://uri" is a URI that defines the given distribution or {\tt externalParameter.}

\section{Flux Balance Constraints}

The SBML ‘Flux Balance Constraints’ ('FBC') package introduced a way to define a system of reactions, each constrained in different ways, together with an objective function to be maximized or minimized.  In this way, some conclusions may be reached about the fluxes in a reaction network without knowing the specifics of the reaction rate equations.  Antimony has now been extended to store this information through the following syntax:

\medskip
\begin{lstlisting}[caption={Flux balance constraint information.}]
// The objective function (for FBC analysis):
OBJF: maximize R16 + R03;

// Constraints:
constraint c0: R16 >= 0
constraint c1: R16 <= 1000
constraint c2: R03 >= -1000
constraint c3: R03 <= 1000
constraint c4: R02 >= -1000
constraint c5: R02 <= 1000

\end{lstlisting}

In the above example, an objective function {\tt OBJF} has been defined that simply states that the sum of the reaction rates of R16 and R03 should be as large as possible (the reaction network itself is not shown here; there are often many reaction defined, none of which have reaction rates).  Some example constraints are listed, collectively stating that reaction R16 must be between 0 and 1000, with R03 and R02 constrained to be between -1000 and 1000.  In a complete model, every reaction would have a lower and upper bound (though it may be positive or negative infinity).

Note that Antimony is currently only compatible with FBC version 1, so models encoded using other versions of the FBC package are translated to version 1 before importing into Antimony.  Some information introduced in later versions of FBC may be lost, though we intend to update FBC support to include that extra information.

\section{Named Stoichiometries}

SBML permits reaction stoichiometries to be set using variables. An example is given below where the second reaction has stoichiometries {\tt n} and {\tt m}. These can be set using either an assignment rule or during initializing. The latter can be useful when comparing how a simulation responds to different stoichiometries. A good example might be investigating how a reaction rate responds to different stoichiometries, such as with a Hill-type mechanism.

Changing the stoichiometry using a simulation by using an assignment rule, is more problematic and depends on how the simulator implements this functionality.  The problem is that the model is being physically changed during the simulation which requires careful management of the differential equation solver. 

\bigskip
\begin{lstlisting}[caption={Setting stoichiometries using variables.}]
  J0: S0 -> S1; k0*S0
  J1: n S1 -> m S2; k1*S1^n

  n := k2/2
  m = 2/3
\end{lstlisting}

\section{Algebraic Rules}

Algebraic rules are defined as equations that are always true but do not declare which variable or variables should be adjusted to ensure that they are true. Algebraic rules have somewhat limited support in some simulators, and are not, for example, supported by Tellurium. They are provided by Antimony solely for users who wish to export the model and use it in other systems that do support such functionality.

To declare an algebraic rule, optionally give it a name, and then declare “0 = [formula]”:

\medskip
\begin{lstlisting}[caption={Examples of algebraic laws using Antimony.}]
0 =  S1*k1 - 10
alg2: 0 = S2*k2 - 20
\end{lstlisting}

\section{Availability}

Antimony is a reusable C/C++ library which means that other applications can host Antimony which will allow the host application to easily import and export SBML via the Antimony language. Source code and binary releases for all platforms are available at~(\url{https://github.com/sys-bio/antimony}). The library itself has a small standard C API making it easy to link to any computer language. For example, Python bindings are available via a pip package~(\url{https://pypi.org/project/antimony/}). A QT-based application called QTAntimony provides a graphical user interface that makes it easy to edit Antimony and translate to and from SBML. Julia bindings~\cite{welsh2023libroadrunner} are available as well as a JavaScript library that can be used from within a web browser~\cite{jardine2023makesbml}.

The source code is licensed under the liberal BSD-3 license, making it suitable for commercial as well as academic use. 

\section{Future Updates}

There are several future updates we wish to implement. The most significant is to implement a human-readable extension for the SBML layout/render packages. Layout~\cite{gauges2006model} permits users to specify the visual layout of a biochemical network model described in the SBML. Render~\cite{bergmann2018sbml} is an extension to the layout package that permits users to specify the look and feel of a laid-out network, in terms of colors, line thickness etc. 

\section*{Acknowledgements}

This work was supported by NIH Biomedical Imaging and Bioengineering award P41 EB023830. The content expressed here is solely the responsibility of the authors and does not necessarily represent the official views of the National Institutes of Health, or the University of Washington. 

\bibliographystyle{apsr}
\bibliography{references}

\end{document}